\documentclass[%
 reprint,
 groupedaddress,
 unsortedaddress,
nofootinbib,
 amsmath,amssymb,
 aps,
 prd,
]{revtex4-1}

\usepackage{graphicx}
\usepackage{subfigure}
\usepackage{dcolumn}
\usepackage{bm}
\usepackage{hyperref}
\usepackage{amssymb}
\usepackage[T1]{fontenc}
\usepackage{ae,aecompl,times}
\usepackage{color}
\usepackage[normalem]{ulem}
\usepackage{amsmath}
\allowdisplaybreaks[4]

\begin{document}

\preprint{}

\title{New method for initial density reconstruction}

\author{Yanlong Shi$^{1,2,}$}
 \email{syl1200@mail.ustc.edu.cn}
\author{Marius Cautun$^{2,}$}
 \email{m.c.cautun@durham.ac.uk}
\author{Baojiu Li$^{2,}$}
 \email{baojiu.li@durham.ac.uk}
 \affiliation{$^1$Department of Astronomy, University of Science and Technology of China, Hefei 230026, Anhui, China\\
 $^2$Institute for Computational Cosmology, Department of Physics, Durham University, Durham DH1 3LE, UK}

\date{\today}

\begin{abstract}
A theoretically interesting and practically important question in cosmology is the reconstruction of the initial density distribution provided a late-time density field. This is a long-standing question with a revived interest recently, especially in the context of optimally extracting the baryonic acoustic oscillation (BAO) signals from observed galaxy distributions. We present a new efficient method to carry out this reconstruction, which is based on numerical solutions to the nonlinear partial differential equation that governs the mapping between the initial Lagrangian and final Eulerian coordinates of particles in evolved density fields. This is motivated by numerical simulations of the quartic Galileon gravity model, which has similar equations that can be solved effectively by multigrid Gauss-Seidel relaxation. The method is based on mass conservation, and does not assume any specific cosmological model. Our test shows that it has a performance comparable to that of state-of-the-art algorithms which were very recently put forward in the literature, with the reconstructed density field over $\sim80\%$ ($50\%$) correlated with the initial condition at $k\lesssim0.6h/{\rm Mpc}$ ($1.0h/{\rm Mpc}$). With an example, we demonstrate that this method can significantly improve the accuracy of BAO reconstruction.
\end{abstract}

\pacs{Valid PACS appear here}
                             

\maketitle


\section{\label{sec:intro}Introduction}

Cosmology used to be a data-starved field, and high-quality observational data only existed in very limited patches of the Universe. The standard approach  for constraining cosmological models was to generate random realisations of our Universe from some theoretical prescription and confront statistical quantities against true data. This situation has dramatically changed since past two decades: from the local to the whole, from the recent to the distant past, a wide array of ongoing and planned astronomical surveys have been making accurate maps of the large-scale matter distribution (e.g., 2{\sc mass} \cite{Skrutskie2006The2MASS}, {\sc wmap} \cite{Hinshaw2012Nine-YearResults}, {\sc planck} \cite{PlanckCollaboration2015PlanckParameters}, {\sc sdss} \cite{Alam2016TheSample}, {\sc des} \cite{TheDarkEnergySurveyCollaboration2005TheSurvey}, {\sc hsc} \cite{Aihara2017TheDesign}, {\sc desi} \cite{DESICollaboration2016TheDesign}, {\sc euclid} \cite{Racca2016TheDesign}, 4{\sc most} \cite{deJong20164MOST:Review}, {\sc lsst} \cite{Ivezic2008LSST:Products}, {\sc tianlai} \cite{Xu2014ForecastsArray}, {\sc ska} \cite{Godfrey2011VeryArray}). The explosion of observational data will rejuvenate cosmology, for example, constrained (in contrast to random) realisations of the Universe can be made from the data available (e.g., \cite{Lavaux,DoumlerConstrainedVelocities,Copi2013Large-AnglePredictions}).

Another possibility enabled by the flooding data is the reconstruction of initial density from an observed distribution of matter, which has both unique theoretical merits and important practical applications. The late-time Universe is a highly complicated entity shaped by various physical processes, such as the nonlinear evolution of large-scale structures under gravity. While these processes enrich the observational features of the Universe, they also make certain targeted features entangled with or contaminated by other effects. One prominent example is the baryonic acoustic oscillations (BAO) scale, a pattern imprinted in the matter distribution by pre-recombination physics which can serve as a standard ruler for measuring the cosmic expansion history. The nonlinear structure growth significantly decreases the BAO signature and therefore weakens its constraining power. If the initial linear density field is recovered by some reconstruction method, it will help enrich the information to be extracted from data in such situations \citep{Eisenstein2007ImprovingPeakb,Padmanabhan:2008dd,Gil-Marin:2015nqa,Seo:2015eyw}.


Reconstruction in the context of cosmology has been visited by various groups which utilise different techniques, for examples, \cite{Peebles1989TracingTime,Nusser1992TracingTime,Croft1996ReconstructionApproximation,Monaco1999ReconstructionCatalogues,Frisch2001ATransportation,Mohayaee2003ReconstructionScheme,Brenier2003ReconstructionProblem,Eisenstein2007ImprovingPeakb,White2015ReconstructionApproximation,Zhu2016NonlinearReconstruction,Pan2016IncreasingReconstruction,Wang2017IsobaricOscillation,Yu2017HaloReconstructionb,Schmittfull2017IterativeReconstruction,Wang2008ReconstructingHalos,Jasche2012BayesianSurveys,Kitaura2012TheSimulations} (see Ref.~\cite{Wang2017IsobaricOscillation,Schmittfull2017IterativeReconstruction} for detailed historical reviews). Ref. \cite{Eisenstein2007ImprovingPeakb}  proposed a simple reconstruction based on Zel'dovich approximation that can sharpen the BAO peak and thus improve the BAO measurement accuracy, which has been demonstrated in real observations \cite{Padmanabhan:2012, Ross:2014qpa, Alam2016TheSample}. This has motivated  many studies of alternative methods of improving the BAO signal \citep[e.g.][]{Tassev2012TowardsOscillations,Zhu2016NonlinearReconstruction,Schmittfull2015EulerianStatistics,Schmittfull2017IterativeReconstruction}. 
Reconstructing the initial conditions helps to reduce the damping of the BAO peaks caused by nonlinear evolution, which, for example, Ref. \cite{Padmanabhan:2008dd} showed in the context of Lagrangian perturbation theory.
The reconstruction methods are not limited to the matter distribution, but they have been extended towards realistic situations, such as using dark matter haloes as tracers and accounting for redshift-space distortions \cite{White2015ReconstructionApproximation,Yu2017HaloReconstruction,Zhu:2017vtj}. 

Recently-proposed iterative methods such as \cite{Zhu2016NonlinearReconstruction,Schmittfull2017IterativeReconstruction} managed to push the scale where linear density information can be reliably recovered to $k\sim0.5$-$0.6h/{\rm Mpc}$, which can lead to a substantial reduction of the uncertainty in BAO measurement \citep{Wang2017IsobaricOscillation,Schmittfull2017IterativeReconstruction}. However, given the importance the BAO reconstruction problem, and that different methods could have different limitations, it will be highly beneficial to develop independent methods which have their own merits.

In this paper, we propose a novel method for initial density reconstruction, which is simple in concept and straightforward in implementation. This method is based on numerical solutions to the Monger-Ampere equation, which originates from mass-conservation and governs the mapping between the initial and final coordinates of some mass distribution. Previous attempts to solve this equation, such as  Ref.~\cite{Frisch2001ATransportation}, reduce this to an optimised mass transportation problem and propose sophisticated optimisation algorithms to solve it by minimising a `cost function' defined by associating the initial and final coordinates of particles. In contrast to this, using the fact that this equation can be recast as a higher-order nonlinear partial differential equation, we propose 
a conceptually straightforward way to solve it using the multigrid relaxation method.

Relaxation algorithms have been  well established as an efficient method to solve elliptical partial differential equations (PDEs), and they are particularly useful for nonlinear PDEs, where the standard fast Fourier transform (FFT) method is of limited use. This algorithm is similar in spirit to the Newton-Ralpson algorithm to solve nonlinear algebraic equations: one starts from an initial guess of the solution, and then iteratively improves the guess until the trial solution is close enough to the true solution. This iterative nature of our method, however, is different from the iterations of other methods, e.g., \cite{Zhu2016NonlinearReconstruction,Schmittfull2017IterativeReconstruction}, in that we do not displace particles in each iteration step, and the iteration here is purely a numerical tool for solution finding. We can move particles to their Lagrangian positions once the solution to the PDE is obtained, although this is unnecessary if we are only interested in having the initial density field. The relaxation method has been used extensively in cosmology, e.g., in $N$-body simulations of standard and non-standard cosmological models: in both cases it is known to have good scaling with parallelisation; we shall illustrate the efficiency of this method using test examples below. Another important property of this new method is that it does not have free parameters, apart from the size of the mesh used to calculate the density field -- there is no need to pre-smooth the density field\footnote{The interpolation scheme to calculate the density field from a discrete set of particles or tracers can be considered as some sort of smoothing, but this method can work with any density assigning scheme including nearest grid point (NGP), clouds in cell (CIC) and triangular-sized clouds (TSC), and some density assignment is unavoidable anyway. Since there is guaranteed to be no shell crossing in this method of reconstruction, it is not necessary to get rid of short-wave modes by additional smoothing.} and all wavelength modes are treated in the same away since the calculation is done purely in real space. Finally, numerical tests show that this method has good convergence properties insensitive to mesh resolution: although our tests in this paper are all done with $N^3$ particles on a mesh with $N^3$ cubic cells, we tried $8N^3$ particles on a $N^3$ mesh, and $N^3$ particles on a $8N^3$ mesh -- in both cases we found similar convergence rates of the relaxation iterations as the default case, and this feature gives the method greater flexibility to deal with various tracer densities.

While our method is different from other state-of-the-art ones such as Ref.~\cite{Zhu2016NonlinearReconstruction,Schmittfull2017IterativeReconstruction}, as we shall show below, it succeeds in recovering the initial density field to the same accuracy as the other methods
which suggests that these methods all face the same limitation: after shell crossing it is no longer possible to uniquely find a particle's Lagrangian position. Our test shows that, despite this limitation, the method can greatly improve the reconstruction of BAO peaks in real space. We will leave extensions to redshift-space reconstruction and biased tracers for future work.

This paper is organised as follows. In Section \ref{sec:method} we describe the main ideas and practical implementations of our method. This shall be followed by some tests of the method and code, and then results showing how well the method works, in Section \ref{sec:results}. We summarise and conclude in Section \ref{sec:summary}. Throughout this paper we use the unit $c=1$, where $c$ is the speed of light, unless otherwise stated.

\section{The method}
\label{sec:method}


To perform reconstruction, we need to link two sets of coordinates: the Lagrangian coordinates $\bf{q}$ and the Eulerian coordinates $\bf{x}$ which correspond to initial and evolved (final) matter distributions separately. There is a unique one-to-one mapping between them before shell crossing starts to take place in structure formation, and the mapping is given by mass conservation,
\begin{eqnarray}\label{eq:mass_conservation}
\rho_{\mathrm{fin}}(\bf{x})\mathrm{d}^3\bf{x}=\rho_{\mathrm{ini}}(\bf{q})\mathrm{d}^3\bf{q},
\end{eqnarray}
where ${\rm d}^3{\bf x}$ and ${\rm d}^3{\bf q}$ are small volume elements in the Eulerian and Lagrangian coordinates, and $\rho_{\rm fin}$ and $\rho_{\rm ini}$ are the densities in those volume elements. The evolution of large-scale structure corresponds to a mapping of ${\bf q}$ into ${\bf x}$, and here we want to solve for the inverse of this mapping, i.e., for given ${\bf x}$ coordinates find the corresponding ${\bf q}$. 
After shell crossing, the mapping is no longer unique, and mass conservation does not guarantee a correct recovery of the initial particle coordinates given their final ones. This highlights the difficulties in reconstructing the initial density field at very small scales where structure formation has been highly nonlinear; however, since we only aim to perform the reconstruction at relatively larger scales, we can still use Eq.~\eqref{eq:mass_conservation}. Indeed, the application of this equation guarantees that no shell crossing happens in the reconstruction process. This is equivalent to assuming that shell crossing has been prevented by a mechanism, similar in spirit to the adhesion model \cite{adhesion,adhesion1,adhesion2,adhesion2b,adhesion3,Hidding2013TheComplexity,Hidding2016TheUniverse}.


As a fine approximation, the initial particle distribution is homogeneous and $\rho_{\rm ini}(\bf{q})=\bar{\rho}$. Defining a displacement potential $\Theta({\bf x})$ so that ${\bf q}=\nabla_{\bf x}\Theta({\bf x})$, Eq.~\eqref{eq:mass_conservation} can be written as
\begin{eqnarray}\label{equ:basic_eqs}
\det\left[\nabla^{i}\nabla_{j}\Theta(\bf{x})\right] &=& \det\left(\frac{\partial q^i}{\partial x^j}\right)\ =\ \frac{\rho_{\textrm{fin}}(\bf{x})}{\bar{\rho}},
\end{eqnarray}
where $i,j=1,2,3$ label the three spatial coordinates and $\det$ denotes matrix determinant. This is a nonlinear mapping which involves matrix operations and therefore is difficult to solve directly, and therefore we shall cast it into a different, easier-to-solve, form.

As mentioned above, 
the objective is to rewrite this equation in the form of a nonlinear elliptical PDE with proper (periodic) boundary conditions, which can be solved using multigrid relaxation. To achieve this, we express $\det\left[\nabla^{i}\nabla_{j}\Theta(\bf{x})\right]$ as a linear combination of $\left(\nabla^2\Theta\right)^3$, $\nabla^i\nabla_j\Theta\nabla^j\nabla_i\Theta\nabla^2\Theta$ and $\nabla^i\nabla_j\Theta\nabla^j\nabla_k\Theta\nabla^k\nabla_i\Theta$, in which Einstein convention for summation is used. After some trivial mathematical calculation, Eq.~\eqref{equ:basic_eqs} becomes: 
\begin{eqnarray}\label{eq:nonlinear_PDE}
&&\frac{1}{6}\left(\nabla^2\Theta\right)^3 - \frac{1}{2}\nabla^i\nabla_j\Theta\nabla^j\nabla_i\Theta\nabla^2\Theta\nonumber\\
&&~~~~~~~~~~~~~~ + \frac{1}{3}\nabla^i\nabla_j\Theta\nabla^j\nabla_k\Theta\nabla^k\nabla_i\Theta\ =\ \frac{\rho_{\rm fin}({\bf x})}{\bar{\rho}}.
\end{eqnarray}
Eq.~\eqref{eq:nonlinear_PDE} looks like a cubic equation for $\nabla^2\Theta$, an observation that is instrumental for our numerical algorithm to work. Of course, this is not entirely true since there are other terms such as $\nabla^i\nabla^j\Theta\nabla_i\nabla_j\Theta$ which depends on $\Theta$: we shall see shortly how to  overcome this hurdle in numerical implementation.

Eq.~\eqref{eq:nonlinear_PDE} is very similar to the field equation in the so-called quartic Galileon model \cite{Nicolis2009GalileonGravity,Deffayet2009CovariantGalileon}, a modified gravity model for which $N$-body simulations have been done in \cite{Li2013SimulatingMeshes} by introducing a multigrid relaxation algorithm to solve the PDE (see Eq.~(32) in \cite{Li2013SimulatingMeshes} for the field equation). In this work, we will follow that method to solve Eq.~\eqref{eq:nonlinear_PDE} in order to tackle the reconstruction problem.

As described in \cite{Li2013SimulatingMeshes}, for numerical reasons it is convenient to split the $\nabla_{i}\nabla_{j}\Theta$ matrix into a diagonal and a traceless part by defining the barred derivatives as
\begin{eqnarray} \label{eq:transformation_derivatives}
\nabla_i\nabla_j\Theta \equiv \frac{1}{3}\delta_{ij}\nabla^2\Theta + \bar{\nabla}_i\bar{\nabla}_j\Theta.
\end{eqnarray}
To appreciate the benefit of this operator splitting, let's recall that, as mentioned after Eq.~\eqref{eq:nonlinear_PDE}, the objective is to rewrite it as a cubic equation for $\nabla^2\Theta$. 
This will enable us to separate the calculation into two steps: (i) solving for $\nabla^2\Theta$ analytically to get $\nabla^2\Theta=\cdots$, and (ii) solving the equation $\nabla^2\Theta=\cdots$ as a linear PDE numerically using relaxation. This means that we use analytical solutions as much as possible, and this has the following advantages: 
\begin{itemize}
\item a linear PDE is in general easier to solve, as it has better convergence properties for relaxation (i.e., the trial guesses can more quickly converge to the true solution);
\item the fact that the PDE we solve takes the form of a cubic equation for $\nabla^2\Theta$ means that there can be multiple solutions for $\nabla^2\Theta$, only one of which can be physical. If we happen to find a wrong branch of solutions, numerically the PDE is satisfied but physically the result will not make sense. We shall see below how, by solving the cubic equation for $\nabla^2\Theta$ analytically, we can ensure that the physical branch of solution is always chosen. 
\end{itemize}
The question now is: how can we make sure that the PDE can be written as a cubic equation for $\nabla^2\Theta$ given that it has other complicated terms containing $\Theta$? This can be seen once we insert Eq.~\eqref{eq:transformation_derivatives} into Eq.~\eqref{eq:nonlinear_PDE}, to obtain
\begin{eqnarray}\label{eq:nonlinear_PDE2}
(\nabla^2\Theta)^3-\frac{9}{2}\bar{\nabla}^i\bar{\nabla}_j\Theta\bar{\nabla}^j\bar{\nabla}_i\Theta\nabla^2\Theta~~~~~~~~~~~~~~~~~~~~~~~~~~~ && \nonumber\\ 
+ 9\bar{\nabla}^i\bar{\nabla}_j\Theta\bar{\nabla}^j\bar{\nabla}_k\Theta\bar{\nabla}^k\bar{\nabla}_i\Theta -27[1+\delta(\bf{x})] &=& 0,~~~
  \label{equ:original_unit}
\end{eqnarray}
where we have introduced the notation of overdensity $\delta(\bf{x})$ as $1+\delta(\bf{x})=\rho_{\textrm{fin}}(\bf{x})/\bar{\rho}$.

The key point here is that we will be solving the PDE on a mesh or, in other words, trying to find the solution $\Theta_{i,j,k}$ for cells labelled by $i,j,k$ (the indices of the cell along the $x,y,z$ directions). In numerical implementations, after performing a second-order-accuracy discretisation, it can be shown that the expression of $\nabla^2\Theta$ depends on $\Theta_{i,j,k}$, while $\bar{\nabla}_i\bar{\nabla}_j\Theta$ does {\it not}\footnote{{For an explicit expression for the discretised $\nabla^2\Theta$, see Eq.~\eqref{eq:discrete_examples}. The explicit expressions for $\bar{\nabla}^i\bar{\nabla}_j\Theta\bar{\nabla}^j\bar{\nabla}_i\Theta$ and $\bar{\nabla}^i\bar{\nabla}_j\Theta\bar{\nabla}^j\bar{\nabla}_k\Theta\bar{\nabla}^k\bar{\nabla}_i\Theta$ can be found from Eqs.~(B1, B3) of Ref.~\cite{Li2013SimulatingMeshes}; they are too lengthy to reproduce here, and so we put them in the Appendix of this paper.}}. This means that, as far as $\Theta_{i,j,k}$ is concerned, Eq.~\eqref{eq:nonlinear_PDE2} can be treated {\it effectively} as a cubic equation for $\nabla^2\Theta$, where the various coefficients of the equation depend only on combinations of $\bar{\nabla}_i\bar{\nabla}_j\Theta$, which, for cell $i,j,k$, do {\it not} depend on $\Theta_{i,j,k}$. This cubic equation can be solved analytically to obtain $\nabla^2\Theta=\cdots$ (the exact expression is rather involved and we present it later). Then, to solve numerically for $\Theta_{i,j,k}$, we insert the discretised expressions into the $\nabla^2\Theta=\cdots$ equation, where the right hand side depends only on $\bar{\nabla}_i\bar{\nabla}_j\Theta$ and thus does not involve directly $\Theta_{i,j,k}$.



In the homogeneous and uniform case ($\delta({\bf x})=0$), we have $\det\left[\nabla^{i}\nabla_{j}\Theta_0({\bf x})\right]=1$, which has an apparent solution
\begin{eqnarray}\label{eq:theta0}
\Theta_0({\bf x}) = \frac{1}{2}(x^2+y^2+z^2),
\end{eqnarray}
corresponding to
\begin{equation}
{\bf q}=\nabla_{\bf{x}}\Theta_0={\bf x}, \nabla^2\Theta_0=3 ~\mathrm{and}~ \bar{\nabla}^i\bar{\nabla}_j\Theta_0=0.
\end{equation}
We can then introduce a new variable $\theta$ as $\Theta\equiv\Theta_0+\theta$ that enables us to rewrite Eq.~\eqref{eq:nonlinear_PDE2} as,
\begin{eqnarray}\label{eq:final_PDE}
\left(\nabla^2\theta+3\right)^3+p\left(\nabla^2\theta+3\right)+q=0, 
\end{eqnarray}
where
\begin{eqnarray}
p & = & -\frac{9}{2}\bar{\nabla}^i\bar{\nabla}_j\theta\bar{\nabla}^j\bar{\nabla}_i\theta, \nonumber \\
q & = & 9\bar{\nabla}^i\bar{\nabla}_j\theta\bar{\nabla}^j\bar{\nabla}_k\theta\bar{\nabla}^k\bar{\nabla}_i\theta-27(1+\delta).
\end{eqnarray}
Note that in the numerical implementation what we solve is the discrete version of Eq.~\eqref{eq:final_PDE}:
\begin{eqnarray}\label{eq:final_PDE2}
\left[\left(\nabla^2\theta\right)_{i,j,k}+3\right]^3+p_{i,j,k}\left[\left(\nabla^2\theta\right)_{i,j,k}+3\right]+q_{i,j,k}=0,\nonumber
\end{eqnarray}
where the subscripts $_{i,j,k}$ means taking the value of a quantity in the cell labelled by $i,j,k$. In particular, $p_{i,j,k}$ and $q_{i,j,k}$ do {\it not} contain $\theta_{i,j,k}$.
Since in a given relaxation iteration for cell $i,j,k$ we only want to find $\theta_{i,j,k}$, in the numerical implementation we can treat Eq.~\eqref{eq:final_PDE} as a cubic equation for $\nabla^2\theta$ as already discussed.

As mentioned above, an advantage of splitting the derivatives into barred and unbarred ones is that one can solve $\nabla^2\Theta$, or rather $\nabla^2\theta$ now, analytically. A cubic equation has three branches of solutions, but not all of which are always real, so we need to decide which of them is physical. This is complicated as the physical solution does not necessarily always stay on the same branch, but varies as coefficients $p, q$ vary.

Defining the discriminant as $$\Delta\equiv\frac{q^2}{4}+\frac{p^3}{27},$$ we can classify the different situations by $\Delta$: 
\begin{itemize}
\item if $\Delta\ge0$, there is only one real root, which must be our physical branch;
\item when $\Delta$ transits across 0 from positive to negative, there are 3 real roots, and the physical one should change continuously.
\end{itemize}
Furthermore, when the density field is homogeneous ($\delta=0$), the solution should be consistent with $\theta=0$ and therefore $\nabla^2\theta=0$. With these constraints, the physical branch of solution is found as
\begin{eqnarray}
\nabla^2\theta &=& -3+\left[-\frac{q}{2}+\Delta^{\frac{1}{2}}\right]^{\frac{1}{3}}+\left[-\frac{q}{2}-\Delta^{\frac{1}{2}}\right]^{\frac{1}{3}},~~{\rm if}~\Delta\geq0;\nonumber\\
\nabla^2\theta &=& -3-\left(-\frac{p}{3}\right)^{\frac{1}{2}}\cos\left[\frac{1}{3}\left(\sigma+2\pi\right)\right],~~{\rm if}~\Delta<0,
\label{equ:root}
\end{eqnarray}
where $\sigma\in[0,\pi]$ is defined by $$\cos\sigma\equiv\frac{3q}{2p}\left(\frac{-3}{p}\right)^{\frac{1}{2}}.$$



The crucial step in our reconstruction algorithm is solving Eq.~\eqref{equ:root} to obtain $\theta({\bf x})$ as well as its gradient. For this purpose we have modified the {\sc ecosmog} code described in \cite{Li2013SimulatingMeshes,Li2012ECOSMOGGravity}, which itself is based on the publicly available $N$-body code {\sc ramses} \cite{Teyssier2002CosmologicalRAMSES}. In the rest of this section we give a brief summary of the algorithm.

\

\subsection{Multigrid Gauss-Seidel relaxation}

{We have mentioned that we will solve Eq.~\eqref{equ:root} using multigrid Gauss-Seidel relaxation. In this subsection we give more details what this amounts to.}

\subsubsection{{Discretisation}} 
          
{Before being able to solve Eq.~\eqref{equ:root} on a mesh, we need to first discretise it. As discussed in passing already, this means replacing the different terms in the equation with their values in mesh cells (labelled by $i,j,k$). The derivatives will then be replaced by finite differences of the values of the quantities in neighbouring cells.}

{One example is the gradient of $\theta$ in the $x$-direction, $\nabla_x\theta$. Knowing the values of $\theta$ in three cells: cell $(i,j,k)$ and its left neighbour $(i-1,j,k)$ and right neighbour $(i+1,j,k)$, this can be calculated using either $$\nabla_x\theta\doteq\frac{1}{h}(\theta_{i+1,j,k}-\theta_{i,j,k}),$$ or $$\nabla_x\theta\doteq\frac{1}{h}(\theta_{i,j,k}-\theta_{i-1,j,k}),$$ where $h$ is the size of the cell. It turns out that these expressions of finite difference lead to a `first-order' accuracy, which means that as we decrease $h$ by using finer cells, the numerical error caused by the discretisation decays linearly with $h$. A scheme with second-order accuracy can be achieved as follows: $$\nabla_x\theta\doteq\frac{1}{2h}(\theta_{i+1,j,k}-\theta_{i-1,j,k}),$$ for which the discretisation error decays as $h^2$ with decreasing $h$. An extension of this to second order derivative $\nabla^2_x\theta$ can be obtained straightforwardly as $$\nabla^2_x\theta\doteq\frac{1}{h^2}(\theta_{i+1,j,k}+\theta_{i-1,j,k}-2\theta_{i,j,k}).$$ This finite difference scheme makes use 3 neighbouring cells, which are said to form a {\it 3-point stencil}. We can use more cells to find expressions of $\nabla^2_x\theta$ with higher-order accuracy, but it is not necessary for this work. 
}

{It can be shown that, up to second-order accuracy, the 3D second-order derivatives $\nabla^2\theta$ and terms like $\left[{\nabla}_i{\nabla_j}\theta\right]_{i\neq j}$ at cell $(i,j,k)$ are given by
\begin{widetext}
\begin{eqnarray}\label{eq:discrete_examples}
\nabla^2\theta &=& \frac{1}{h^2}(\theta_{i+1,j,k}+\theta_{i-1,j,k}+\theta_{i,j+1,k}+\theta_{i,j-1,k}+\theta_{i,j,k+1}+\theta_{i,j,k-1}-6\theta_{i,j,k} ) + o(h^2);\\
{\nabla}_x{\nabla_y}\theta &=& \frac{1}{4h^2}(\theta_{i+1,j+1,k}+\theta_{i-1,j-1,k}-\theta_{i-1,j+1,k}-\theta_{i+1,j-1,k}) + o(h^2);\\
{\nabla}_y{\nabla_z}\theta &=& \frac{1}{4h^2}(\theta_{i,j+1,k+1}+\theta_{i,j-1,k-1}-\theta_{i,j-1,k+1}-\theta_{i,j+1,k-1}) + o(h^2);\\
{\nabla}_x{\nabla_z}\theta &=& \frac{1}{4h^2}(\theta_{i+1,j,k+1}+\theta_{i-1,j,k-1}-\theta_{i-1,j,k+1}-\theta_{i+1,j,k-1}) + o(h^2),
\end{eqnarray}
\end{widetext}
where $o(h^2)$ is a shorthand notation for all higher-order contributions.
Interested readers may check Appendix \ref{sec:appendixa} for expressions of more complicated quantities. After discretisation, Eq.~\eqref{equ:root} becomes:
\begin{widetext}
\begin{eqnarray}\label{equ:discretisition}
\mathcal{L}^h\left[\theta_{i,j,k}\right] &=& \frac{1}{h^2}(\theta_{i+1,j,k}+\theta_{i-1,j,k}+\theta_{i,j+1,k}+\theta_{i,j-1,k}+\theta_{i,j,k+1}+\theta_{i,j,k-1}-6\theta_{i,j,k} )-\Sigma_{i,j,k} = 0,
\end{eqnarray}
\end{widetext}
where $\Sigma_{i,j,k}$ is right-hand side of Eq.~\ref{equ:root}, which is a function of $\bar{\nabla}^i\bar{\nabla}_j\theta\bar{\nabla}^j\bar{\nabla}_i\theta$, $\bar{\nabla}^i\bar{\nabla}_j\theta\bar{\nabla}^j\bar{\nabla}_k\theta\bar{\nabla}^k\bar{\nabla}_i\theta$ and $\delta$, all evaluated in cell $(i,j,k)$ and all independent of $\theta_{i,j,k}$. The superscript $^h$ in $\mathcal{L}^h$ reminds us that $\mathcal{L}^h$ is the differential operator on a mesh of cell size $h$.}

{When implemented into the {\sc ecosmog} code, Eq.~\eqref{equ:discretisition} and all terms in it are actually expressed using internal code unit. In our code the internal units are specified by using the tildered dimensionless quantities instead of the untilered dimensional quantities as follows:
\begin{eqnarray}
	\tilde{x}=\frac{x}{B},\qquad\tilde{\rho}=\frac{\rho a^3}{\rho_c\Omega_m},\qquad\tilde{\theta}=\frac{\theta}{B^2},
\end{eqnarray}
where $B$ is the comoving size of the simulation box and $\rho_c\Omega_m$ is the mean matter density.}


\subsubsection{{Gauss-Seidel relaxation}}

{
As briefly described in Section \ref{sec:intro}, the relaxation method is a method to update the initial guess of the solution $\theta_{i,j,k}$ for all $(i,j,k)$ iteratively until the trial solution becomes sufficiently close to the true solution.} 

{The explicit iteration scheme is 
\begin{eqnarray}\label{eq:update}
\theta^{\mathrm{new}}_{i,j,k} &=& \theta^{\mathrm{old}}_{i,j,k}-\frac{\mathcal{L}^h\left[\theta^{\mathrm{old}}_{i,j,k}\right]}{\partial \mathcal{L}^h\left[\theta^{\mathrm{old}}_{i,j,k}\right]/\partial \theta^{\mathrm{old}}_{i,j,k}},
\end{eqnarray}
in which $\theta^{\rm old}_{i,j,k}$ is the value of $\theta$ in cell $(i,j,k)$ at the present iteration (or the initial guess if this is the first iteration), while $\theta^{\rm new}_{i,j,k}$ is the value of $\theta$ for the same cell at the next iteration.}

{For this relaxation scheme to work, the discrete PDE has to be supplemented by an initial guess and a suitable boundary condition. An advantage of rewriting the original PDE (for $\Theta$) in terms of $\theta$ is that it makes it easier to write down an initial guess for {\it all} cells: $\theta_{i,j,k}=0$. This is a simplification because there are a huge number of cells in the computation, and it is generally more difficult to motivate an initial guess which differs cell by cell than simply using $0$ in every cell. The use of $\theta$ instead of $\Theta$ also makes it easier to set up periodic boundary conditions for the relaxation: this is because, according to Eq.~\eqref{eq:theta0}, even in the case of a homogeneous density field where $\Theta=\Theta_0$, $\Theta$ does not satisfy a periodic boundary condition}.

{The partial derivative with respect to $\theta_{i,j,k}$ in Eq.~\eqref{eq:update} is evaluated at the {\it present} iteration, and according to Eq.~\eqref{equ:discretisition} it is given by the simple form $$\partial \mathcal{L}^h\left[\theta^{\mathrm{old}}_{i,j,k}\right]/\partial\theta^{\mathrm{old}}_{i,j,k}=-\frac{6}{h^2},$$ thanks to the facts that we have solved the cubic equation to find a linear equation for $\nabla^2\theta$ whose right-hand side does not contain $\theta_{i,j,k}$.}

{In practice, there are a large number of cells for {\it all} of which the value of $\theta_{i,j,k}$ is updated during this iteration process. The updates can be arranged in different ways. For example, since the discretised operator $\mathcal{L}^h\left[\theta_{i,j,k}\right]$ in Eq.~\eqref{equ:discretisition} depends on not only $\theta_{i,j,k}$ and $\delta_{i,j,k}$, but the values of $\theta$ in neighbouring cells such as $\theta_{i\pm1,j\pm1,k\pm1}$, when the update in Eq.~\eqref{eq:update} is carried out for cell $(i,j,k)$, the $\theta$ values in some of its neighbouring cells may have already been updated. It is certainly possible to choose to {\it not} use these updated neighbour-cell values of $\theta$ in Eq.~\eqref{eq:update}, such as the {\it Jacobi} method. In our implementation, however, we use the {\it Gauss-Seidel} method, where the updated neighbour-cell values of $\theta$ {\it are} used in Eq.~\eqref{eq:update} as soon as they are available.}

{The process during which all cells have their $\theta$ values updated is called a sweep. During one sweep one can in practice choose different orders to update the cells, and in our code we use the so-called black-red chessboard ordering. It is helpful to visualise this using a chessboard where cells which are direct neighbours of each other (i.e., they have a common face) have different colours (black vs red), while cells which are diagonal neighbours have the same colour. The iteration sweep is divided into two sub-sweeps, during which only the red and the black cells get updated each time respectively. {We notice that this order is not particularly consistent with the way we discretise our equation: $\nabla^2\theta$ depends only on the direct (i.e., different-colour) neighbours of cell $(i,j,k)$, while 
$\Sigma_{i,j,k}$ depends only on the diagonal (i.e., same-colour) neighbours. As the latter is used as the source of the equation, this means that within a given subsweep the cells whose $\theta$ values are used to calculate the sources are constantly updated -- this is different from the standard Poisson equation, for which the source does not depend on the $\theta$ value of any cell and so stays unchanged for a {full} sweep: this is why solving our nonlinear PDE is less efficient than solving the Poisson equation\footnote{{Naively, one would expect that, if the source keeps changing after updating every cell, then it is more difficult for the relaxation to converge, because {\it the equation itself} keeps changing. This is why, even though our equation has been rewritten in the form of a standard Poisson equation, $\nabla^2\theta=\cdots$, the relaxation converges more slowly than it does for the standard Poisson equation}.}. One possible way to improve is to use more complex ordering schemes to do the sweep, for example by separating the sweep across the simulation mesh into 4 (rather than 2) subsweeps. We do not pursue those possibilities in this work because the black-red scheme works reasonably well for our reconstruction problems}.}

{To check if the trial solution after an iteration step has become sufficiently close to the true solution, we use the residual $\epsilon$ defined as $$\epsilon\equiv\left[\frac{1}{N^3}\sum_{i,j,k}\left(\mathcal{L}^h[\theta_{i,j,k}]\right)^2\right]^{1/2},$$ where the summation is over $N^3$ cells in the mesh. Evidently, if the trial solution is exactly equal to the true solution for all cells, then $\epsilon=0$. In general, there is always numerical error so that $\mathcal{L}^h\left[\theta_{i,j,k}\right]\neq0$ (which is why Eq.~\eqref{eq:update} makes sense!), but if the algorithm is stable then $\epsilon$ decreases with more iterations. In our code we set a criterion that if $\epsilon<10^{-8}$ the relaxtion is deemed to be {\it converged} and the iteration stops.}

\subsubsection{{Multigrid V-cycles}}

{The purpose of relaxation iterations is to reduce the error of the trial solution. For the error wave modes that are similar in size to the cell spacing, $h$, this is usually achieved relatively quickly, after a small number of iterations (depending on the nonlinearity of the PDE being solved). Qualitatively, this is as expected since each iteration uses only the nearest neighbours to update the trial solution of cell $(i,j,k)$. Decreasing the long error wave modes generally takes many more iterations, and hence much longer computational time, posing a challenge to the efficiency of the relaxation method.}

{In practical implementations, a speed-up of the convergence rate is often achieved using the so-called multigrid method. Here, after a few iterations on level $h$ (we use the cell size $h$ to label the level of the mesh because multigrid methods use more than one mesh as we will describe now), when the error wave modes comparable to $h$ have been reduced and the convergence starts to slow down due to the inefficient reduction of long wave modes of the error, one moves the equation to a coarser mesh with cell size $H=2h$ (labelled as level $H$). The idea is that by using a second mesh with larger cell size, the wave modes comparable to $H$ will be reduced more quickly, therefore improving the convergence rate.} 

{The coarsification of the discrete PDE from level $h$ to level $H$ is done using the so-called {\it restriction} operator $\mathcal{R}$. Suppose that the solution at level $h$ is $\hat{\theta}^h$ before moving to level $H$, and that numerically $\hat{\theta}^h$ satisfies
\begin{eqnarray}\label{eq:PDE_h_w_error}
\mathcal{L}^h\left[\hat{\theta}^h_{i,j,k}\right] &=& d^h_{i,j,k},
\end{eqnarray}
where $d^h$ is the remaining error on level $h$ (it should be zero or nearly zero if the solution is accurate), then the PDE to be solved on level $H$ is
\begin{eqnarray}\label{eq:PDE_H}
\mathcal{L}^H\left[{\theta}^H_{i,j,k}\right] &=& \mathcal{L}^h\left[\mathcal{R}\hat{\theta}^h_{i,j,k}\right] - \mathcal{R}d^h_{i,j,k},
\end{eqnarray}
which is a coarsified version of 
\begin{eqnarray}
\mathcal{L}^h\left[{\theta}^h_{i,j,k}\right] - \mathcal{L}^h\left[\hat{\theta}^h_{i,j,k}\right] = -d^h_{i,j,k},\nonumber
\end{eqnarray}
which itself is the difference between Eq.~\eqref{equ:discretisition} and Eq.~\eqref{eq:PDE_h_w_error}. Eq.~\eqref{eq:PDE_H} is then solved using a similar Gauss-Seidel relaxation on level $H$ to find the (approximate) solution $\hat{\theta}^H$, and the old approximate solution on level $h$, $\hat{\theta}^h$, can be corrected as
\begin{eqnarray}\label{eq:corrections}
\hat{\theta}^{h,{\rm new}} &=& \hat{\theta}^h + \mathcal{P}\left(\hat{\theta}^H-\mathcal{R}\hat{\theta}^h\right),
\end{eqnarray}
where $\mathcal{P}$ is the so-called {\it prolongation} operator. $\mathcal{R}$ and $\mathcal{P}$ are responsible for the forward and backward interpolations between the fine ($h$) and coarse ($H$) levels, and they can be defined in different ways in practice. As an example, in 3 dimensions each coarse cell covers 8 fine (son) cells, and in the $\mathcal{R}$ operation the value of a quantity in a coarse cell can be taken as the average of its values in the 8 son cells.}

{It should be clear that the principle can be applied to use further coarser meshes to speed up the reduction of longer wave modes of the error, and this use of multiple grids is why the method is called {\it multigrid} relaxation. In our implementation, we use a hierarchy of meshes with the coarsest one having $4^3$ cells. The code does restrictions consecutively from the finest mesh to the coarsest one, solving for $\theta^h$ on all levels, and then does prolongations all the way back to the finest level to correct the solution there using Eq.~\eqref{eq:corrections}. Such an arrangement of going forward and backward across the meshes is intuitively called a {\it V-cycle}.}


\subsection{{Initial density reconstruction}}

{Once $\theta$ and hence $\Theta$ has been obtained on the whole computational mesh, it is straightforward to reconstruct the initial density field from that. In practice this consists of the following steps:}

{\it $\bullet$ Step B1: Finding iso-$q$ lines}: the code outputs $\theta({\bf x})$ and its gradient $\nabla_{\bf x}\theta({\bf x})$ on a regular $\bf{x}$ grid; from this we compute lines of equal $q_{x}, q_{y}$, $q_{z}$ coordinates (called iso-$q$ lines).

{\it $\bullet$ Step B2: Identifying the displacement field $\chi({\bf q})$}: we want the displacement field ${\bf\chi}({\bf q})$ given by $\bf{\chi}=\bf{q}-\bf{x}$, as a function of ${\bf q}$, which can be done once we have the iso-${q}$ lines equally spaced in ${\bf q}$ (from the process to find the iso-$q$ lines, we know the ${\bf x}$ coordinates of the ${\bf q}$ grids).

{\it $\bullet$ Step B3: Calculating the reconstructed density field}: this can be obtained by taking the divergence wrt ${\bf q}$,
\begin{equation} \label{eq:reconstructed_density}
\delta_r\ =\ \nabla_{\bf{q}}\cdot\bf{\chi}.
\end{equation}
This is calculated by the publicly-available {\sc dtfe} code \cite{cautun2011dtfe}, which is based on Delaunay tessellation.

{The recovered density field is, to leading order approximation, the initial density linearly extrapolated to the redshift of the reconstruction. To see this, we write the reconstructed displacement field as $\chi\equiv\chi_{\rm Z} + \chi_{\rm corr}$, where $\chi_{\rm Z}$ is the first-order contribution (the Zel'dovich approximation) and $\chi_{\rm corr}$ denotes higher-order and methodological corrections. Inserting this into Eq.~\eqref{eq:reconstructed_density}, which defines the reconstructed density, results in 
\begin{equation}\label{eq:delta_r_Z}
\delta_r\ =\ D_+ \delta_{\rm ini} + \nabla_{\bf{q}}\cdot \chi_{\rm corr},
\end{equation}
where $D_+$ is the linear growth factor at the redshift at which the reconstruction is done and $\delta_{\rm ini}$ the initial density field. The second term represents corrections due to the facts that on the scale corresponding to the reconstruction grid cell size the growth of structures might have progressed pass the first order Zel'dovich approximation, and that we are treating a realistic, shell-crossed, particle distribution as if shell crossing had not happened. Choosing a coarser reconstruction grid decreases the amount of nonlinear evolution and thus reduces this correction term, but only at the expense of recovering the density at fewer locations (as there are fewer cells) and so potentially losing some useful information.
}

\begin{figure}[htbp]
\centering
\includegraphics[width=0.52\textwidth]{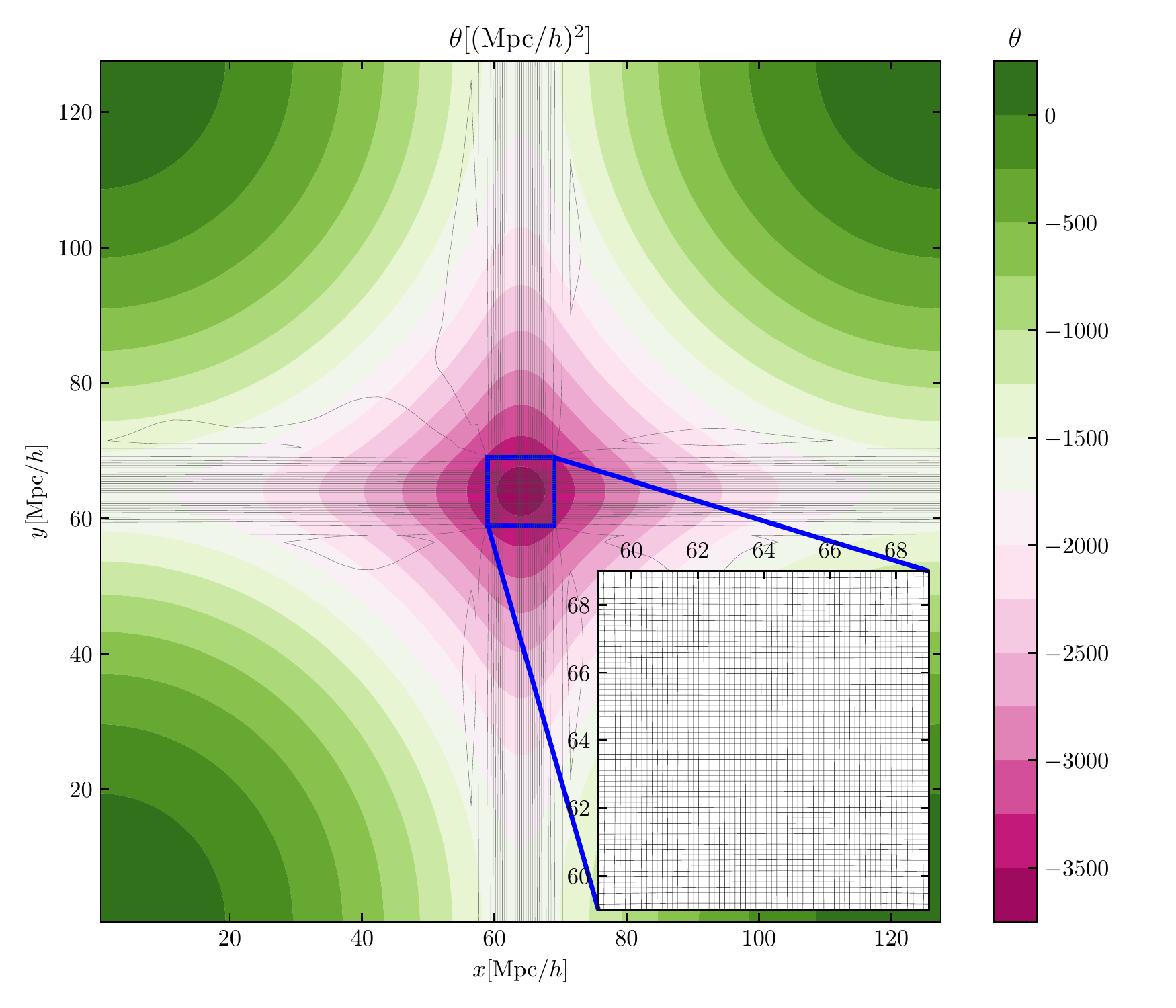}
\caption{(Colour Online) Method test using a cubic particle distribution. The final density field is a uniform distribution of $128^3$ particles inside a $10\mathrm{Mpc}/h$-per-side cubic box placed at the centre of the simulation box of size $128\mathrm{Mpc}/h$, with the faces of the two boxes parallel to each other. The colour map is the $\theta$ field computed by the new reconstruction method, with the colour bar on the right indicating the values of $\theta$. The horizontal and vertical gray lines are respectively lines with equal Lagrangian $q_y$ and $q_x$ coordinates. Note that only gray lines within the central blue box (which is zoomed in in the lower right corner of the figure) are meaningful -- particles at the corners of this blue box are at the corners of the simulation box in the initial (reconstructed) distribution. As mentioned in the abstract, the reconstruction method is based solely on mass conservation, with no information about how the density field has evolved. Only $64\times64$ (out of $128\times128$) lines are shown here for a clear view, and the plot is a $0.64\mathrm{Mpc}/h$-thick slice perpendicular to the $z$-axis at the middle of the box. Note that the $q$-grid here is regularly spaced except the region near the edge, where numerical errors occurred.}
\label{fig:cubic}
\end{figure}

\begin{figure}[htbp]
\centering
\includegraphics[width=0.52\textwidth]{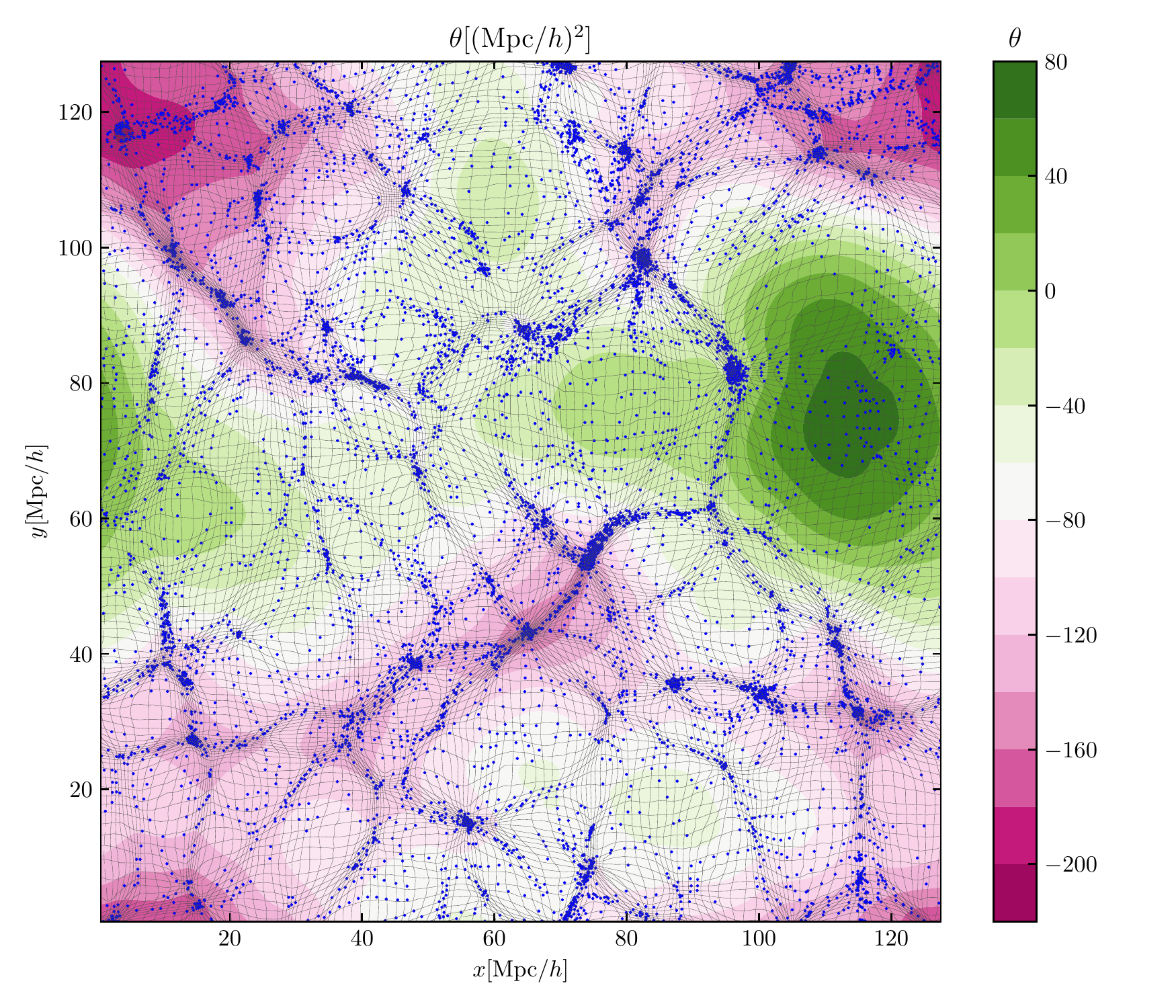}
\caption{(Colour Online) Similar to Figure \ref{fig:cubic}, but now the reconstruction is performed starting from a proper $N$-body simulation snapshot at $z=0$. 
The blue dots are the particles inside the slice shown at their Eulerian coordinates at $z=0$. Note how the $q$-grid distorts following the distribution of matter particles and becomes concentrated (expanded) in overdense (underdense) regions.
}
\label{fig:grids}
\end{figure}

\section{Code test, results and performance}
\label{sec:results}

\subsection{\label{sec:testing}Cubic Test}

Before applying the code to real reconstruction problems, we present a test
using a simple but non-trivial configuration following \cite{Li2013SimulatingMeshes}. {Such tests are important because they serve as useful sanity checks of the code and the algorithm, as well as helping us to build up intuition from simplified problems.}

As we use a cubic box with periodic boundary conditions, an ideal test -- to preserve the symmetry -- is to have the particles distributed in a smaller cubic volume at the centre of the simulation box. We do this by uniformly distributing the particles in this small region, and  call it the `cubic test'. {Should it work properly, the reconstruction is expected to move the particles and uniformly fill the whole box, because the `initial' particle distribution (in ${\bf q}$ coordinate) is uniform. }

In practice, we sampled $128^3$ static particles and put them in a small cubic region with $10\mathrm{Mpc}/h$ per side, while the full box size is 128$\mathrm{Mpc}/h$. Figure \ref{fig:cubic} displays a slice through the simulation box near the centre. Here we've plotted iso-$q_x$ and iso-$q_y$ contours to represent $\bf{q}$ grids, and as expected they are uniform inside the small cubic region at the box centre. {This is because, when particles move, they carry their Lagrangian coordinates with them. Given that there is no shell crossing, and that the particle distribution is uniform both in the initial state (when they fill the whole simulation box) and in the final state (the state on which the reconstruction is done, when the particles fill the central cubic subbox), particles with the same $x$ ($y$ or $z$) coordinates in the final state should have the same $q_x$ ($q_y$ or $q_z$) Lagrangian coordinates, meaning that the iso-$q$ lines must form a uniform grid inside the central cubic region, which our test successfully confirms.} 

Note that on the edges of this cubic region, the density field has a sharp jump, causing slightly larger errors in our numerical solutions, which is why the iso-$q$ grid is less uniform there. {Cosmological distributions do not have such sharp unphysical jumps, thus this limitation of the method is unimportant when reconstructing the cosmological density field.}


\subsection{\label{sec:vis}Visual check of a real construction problem}

Having verified that the code works properly, we then applied it to reconstruction of initial conditions for a $z=0$ particle distribution produced using an $N$-body simulation. The simulation was carried out by {\sc ramses} and followed $128^3$ particles in a cubic box of length $128$ $h^{-1}\mathrm{Mpc}$ from $z=49$ to $z=0$, from an initial condition generated using second order Lagrangian perturbation theory ({\sc 2lptic}\cite{Crocce2006TransientsSimulations}). We then performed the reconstruction with the $z=0$ snapshot.

\begin{figure}[htbp]
	\centering
	\includegraphics[width=0.505\textwidth]{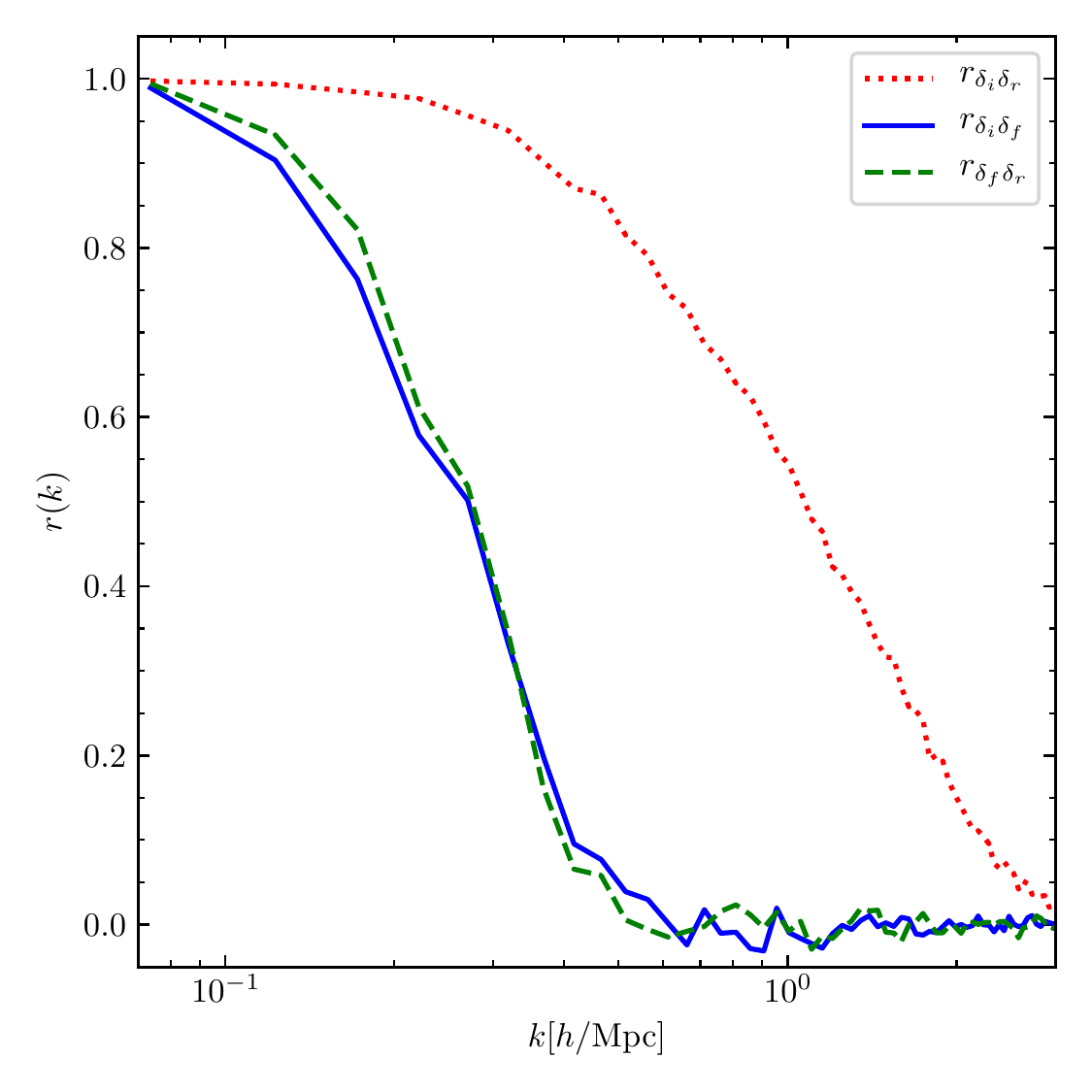}
	\caption{(Colour Online) The correlation coefficient $r(k)$ between the initial and final (green solid line), initial and reconstructed (red dotted), and final and reconstructed (blue dashed) density fields, which show that the reconstruction successfully recovers information in the initial density field that gets lost due to structure formation.}
	\label{fig:correlation}
\end{figure}

Figure \ref{fig:grids} displays the reconstruction result for a thin slice through the simulation box, with the iso-$q_x$ and iso-$q_y$ grids shown as grey lines, where we have also overplotted the particles in this slice (blue dots) and the $\theta$ field (the coloured map). As expected, this gives a distorted $\bf{q}$ grid, where the grid tends to shrink for high-density regions  (which also aligns with filaments), while expand for low-density regions. 

{Again, the distortion of the iso-$q$ grids can be understood as the consequence of particles carrying their Lagrangian coordinates while clustering. In the initial condition of the simulation, the particles are on an almost uniform initial grid of Lagrangian coordinates; when they form clusters and filaments, the initially uniformly-spaced grid lines concentrate, leading to the distortions well aligned with the filaments. In low density regions, particles flow apart and lead to iso-$q$ grids that are further apart and potentially distorted by the large scale tidal field.}

{The $\theta$ field, on the other hand, is effectively the `potential' of the displacement field: ${\bf x}-{\bf q}=-\nabla_{\bf x}\theta({\bf x})$. In low-density regions where particles evacuate from, $\theta$ reaches a local maximum; while in high-density regions particles fall into, $\theta$ reaches a local minimum, in an analogy to the Newtonian potential. This is consistent with what Figs \ref{fig:cubic} and \ref{fig:grids} show.}


\subsection{\label{sec:cor}Quantitative checks of reconstruction}

To go beyond the qualitative visual inspections and check the performances of the method quantitatively, we have measured the auto and cross matter power spectra of the initial, final and reconstructed density fields, and checked that the auto power spectra of the initial and reconstructed density fields have similar shapes down to $k\approx0.5h/\mathrm{Mpc}$. 

\begin{figure}[htbp]
\centering
\includegraphics[width=0.505\textwidth]{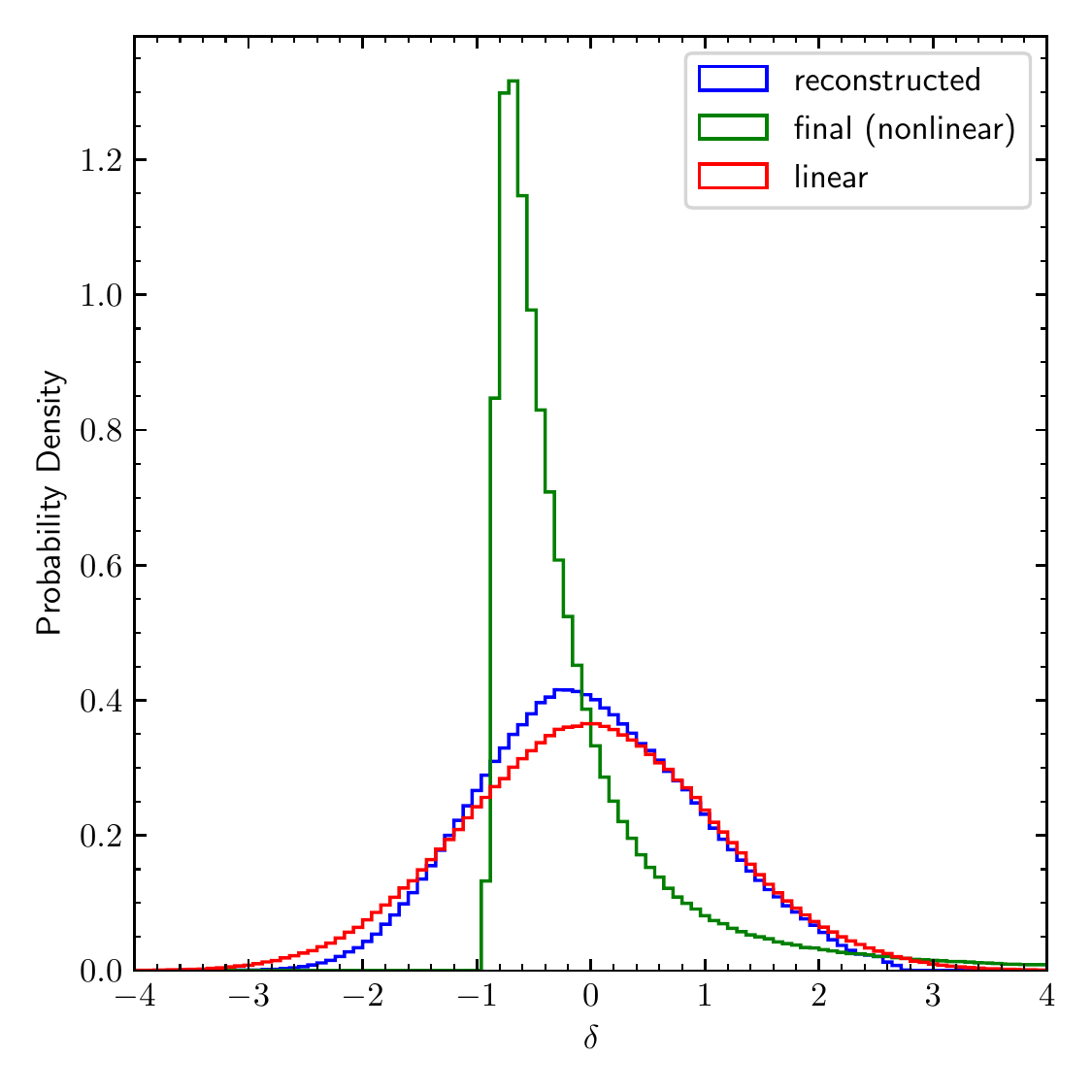}
\caption{(Colour Online) Normalised histograms of the initial ($z=49$; red), final ($z=0$; green) and reconstructed (blue) density field values, and $\delta=\rho/\rho_0-1$ is the overdensity. The initial density field has been linearly extrapolated to $z=0$ by multiplying with the linear growth factor $D_+\approx35.8$. All density fields have been smoothed with a $2$ $h^{-1}\mathrm{Mpc}$ Gaussian filter.}
\label{fig:rhobin}
\end{figure}

To quantify how much of the information in the initial density field has been successfully recovered by the reconstruction, we define the correlation coefficient of two density fields $\delta_1$ and $\delta_2$, {denoted by $r_{\delta_1\delta_2}$}, as
\begin{eqnarray}
r_{\delta_1\delta_2}=\frac{\tilde{\delta}_1\tilde{\delta}_2^\ast+\tilde{\delta}_1^\ast\tilde{\delta}_2}{2\sqrt{\tilde{\delta}_1\tilde{\delta}_1^\ast}\sqrt{\tilde{\delta}_2\tilde{\delta}_2^\ast}},
\end{eqnarray}
where $\tilde{\delta}$ is Fourier transform of $\delta$, and $^\ast$ indicates the complex conjugate.

In Figure \ref{fig:correlation} we show the correlations among the initial ($\delta_i$), final ($\delta_f$) and reconstructed ($\delta_r$) density fields. While the correlation between $\delta_i$ and $\delta_f$ starts to fade at $k\approx0.07h/\mathrm{Mpc}$, the correlation between $\delta_i$ and $\delta_r$ remains substantial even on relatively small scales: {$r_{\delta_i\delta_r}\geq$ $0.95$, $0.85$, $0.8$ and $0.55$ respectively at $k/(h{\rm Mpc}^{-1})\approx$ $0.3$, $0.5$, $0.6$ and $1.0$. Comparing to previous works, a quick on-screen measurement shows that the values of $r_{\delta_i\delta_r}$ at these scales are $0.95$, $0.85$, $0.78$ and $0.52$ for the $\mathcal{O}(1)$ reconstruction and $0.96$, $0.86$, $0.8$ and $0.55$ for the $\mathcal{O}(2)$ reconstruction method in \citep[][Fig.~4]{Schmittfull2017IterativeReconstruction}, and 
$0.95$, $0.8$, $0.7$, $0.45$ 
in \cite[][Fig.~3]{Zhu2016NonlinearReconstruction}, although we remark that these comparisons are only indicative because of the possibly different simulation specifications, reconstruction settings and correlation measurements. A fairer comparison can be made by running all these methods on the same particle snapshot, a possibility that we leave for future. Hence, we conclude that the performances of these methods are broadly consistent.
}


In Figure \ref{fig:rhobin} we show the histograms of the density values for $\delta_i, \delta_f$ and $\delta_r$, which have been normed to unity. All three density fields have been smoothed by a Gaussian window function of width $2$ $h^{-1}\mathrm{Mpc}$, and we have extrapolated the initial density field by multiplying $\delta_i$ with the linear growth factor $D_+(z=0)$ ({cf.~Eq.~\eqref{eq:delta_r_Z}}; $35.82$ for our chosen cosmology). 

As expected, the nonlinear density field $\delta_f$ is {strongly} non-Gaussian with a sharp cut-off at $\delta_f=-1$ and a long tail at positive $\delta_f$. On the other hand, the reconstructed and (linearly extrapolated) initial density fields have similar distributions, both following a Gaussian shape of similar widths and peak positions (with the one for $\delta_r$ slightly skewed). The Gaussianisation of the reconstructed density field is another indicator that the new method works well. Note also that, because there is no shell crossing in this reconstruction, the iso-${\bf q}$ curves do not intersect, so $\nabla_{\bf q}\cdot{\bf \chi} = 3 - \nabla_{\bf q}\cdot{\bf x}<3$, which explains why the reconstructed densities do not go beyond $\delta_r=3$ \cite{Neyrinck,Zhu2016NonlinearReconstruction}. {On the other hand, both $\delta_r$ and $\delta_i$ have a long tail at $\delta<-1$: the value is not bound by $\delta=-1$ because to leading order $\delta_r$ is the same as $D_+\delta_i$, c.f.~Eq.~\eqref{eq:delta_r_Z}; though $|\delta_i|\ll1$ in general, the multipliation by $D_+\gg1$ can cause a negative $\delta_i$ to go below $-1$).}

We have also compared the initial and reconstructed density maps visually, and confirmed that they resemble each other closely. {Furthermore, we have tried the 1-point Gaussianisation technique \cite{Weinberg1992,Neyrinck2009} to make the reconstructed density field perfectly Gaussian, but this indeed slightly decreases the correlation between it and the initial density field.}

\subsection{\label{sec:bao_rec}{Application to BAO reconstruction}}

\begin{figure}[htbp]
\centering
\includegraphics[width=0.48\textwidth]{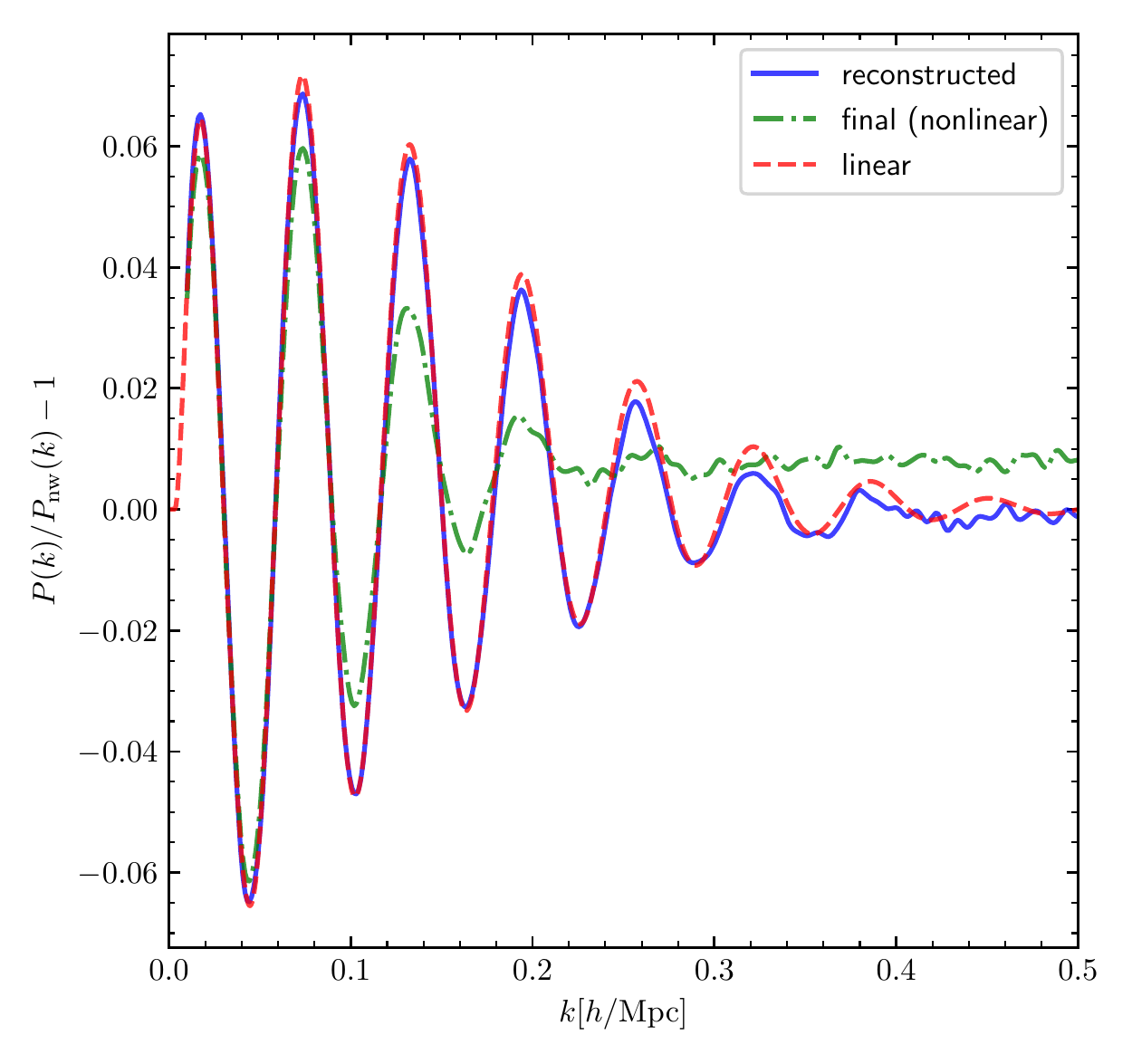}
\caption{{(Colour Online) Comparison of the BAO signals from the initial (red dashed line), final (green dot-dashed) and reconstructed (blue solid) density fields. 
As we can see, for the reconstructed field the amplitude and positions of the peaks are in good agreement with those of the initial linear density field even after $k\approx 0.3 h/\mathrm{Mpc}$, while in the nonlinearly evolved density field the peak features are degraded substantially. This demonstrates that the reconstruction can greatly improve the accuracy of BAO measurements. 
}}
\label{fig:bao}
\end{figure}

{As mentioned in the introduction, a main motivation of reconstruction in modern cosmology is to improve the recovery of BAO features. This is illustrated in Figure \ref{fig:bao}, where we compare the BAO signal in the initial conditions, and the one at $z=0$ calculated from the evolved density distribution and from the reconstructed one.}

{
For this test we used two simulations of the same $\Lambda$CDM cosmology, with $512^3$ particles in a $1$ $h^{-1}{\rm Gpc}$ box starting from initial conditions generated using the same phases, but one with BAO wiggles in the input linear power spectrum and the other without. The grid size used for the reconstruction is $512^3$. To illustrate the BAO feature, we calculate the quantity, $P(k)/P_{\mathrm{nw}}(k)-1$, where $P_{\mathrm{nw}}(k)$ indicates a non-wiggle template which was generated by an initial condition without the BAO signal.} 

{As one can see from Fig.~\ref{fig:bao}, in the nonlinearly evolved density field, the high-$k$ peaks are both weakened and shifted, degrading the BAO signal; the decrease of the BAO signal starts even at $k\sim0.07h{\rm Mpc}^{-1}$ and the peaks become invisible at $k\geq0.2 h\mathrm{Mpc}^{-1}$. However, the reconstructed density field has BAO features that agree very well with the linear density field even after $k=0.3$ $h\mathrm{Mpc}^{-1}$, and the peaks are still visible after that, such as at $k\sim0.4$ $h{\rm Mpc}^{-1}$.}

\section{Discussion and Summary}
\label{sec:summary}

The reconstruction of initial density fields from observational data is a long-standing problem in cosmology and has attracted considerable interests recently. In this work we have proposed a new efficient method to do this by solving the PDE governing the mapping between the Lagrangian and Eulerian coordinates of particles, and implemented it numerically. 

We have carried out a range of visual and quantitative tests of the new method (and the code), to check that it works well. In particular, the reconstruction removes a large part of the non-Gaussianity in the density distribution that has been produced by the nonlinear evolution of large-scale structures, and successfully restores information {present} in the initial density field that is not readily available in the final evolved density field. This can be observed by the fact that {the correlation between initial and reconstructed density fields} remains $\gtrsim 0.8$ down to scales of $k\sim0.6h/\mathrm{Mpc}$. These performances are comparable to those of some other algorithms proposed very recently \cite{Zhu2016NonlinearReconstruction,Schmittfull2017IterativeReconstruction} {(see Section \ref{sec:cor} for an explicit comparison)}. {The reconstruction leads to a significant improvement of BAO signal down to scales, $k\approx 0.4~h/\mathrm{Mpc}$, as can be assessed from Fig.~\ref{fig:bao}.}

The new method {calculates the displacement field by using multigrid relaxation, which is instrumental for fast and efficient convergence. 
For the reconstruction shown here, we achieve convergence} in eight V-cycles (each V-cycle consists of about $30$ relaxation iterations), taking less than one minute on 64 CPUs (
we also tested it for a larger reconstruction problem, using $512^3$ particles on a grid of size $512^3$ in a simulation box of size $500{\rm Mpc}/h$, and it also converged in eight V-cycles, taking about six minutes on 360 CPUs). Intuitively, this is understandable: though Eq.~\eqref{eq:final_PDE} appears local, mass conservation is global and the algorithm needs to walk through the entire simulation box to decide the coordinate mapping; multigrid, by solving the equation on hierarchically coarser grids, allows increasingly bigger steps for the `walk' by which it speeds up the rate of convergence to the final solution.

We expect to find use of this algorithm in various applications, such as the removal of nonlinear evolution contamination in measurements of the BAO peak, the precision requirement of which has greatly increased with upcoming galaxy surveys such as {\sc desi} and {\sc euclid}. Another potential development is to {use more realistic tracers, e.g., different populations of galaxies, of the dark matter field for the reconstruction and} include the redshift space distortion effect in the reconstruction process, so that the latter can be done for observed galaxy catalogues where galaxy positions are given in redshift space. The accurate mapping between Lagrangian and Eulerian coordinates will also allow to infer the initial density field from the observed cosmic web, and therefore understand evolution of structures such as cosmic voids. We will leave the investigation of these possibilities to future work.

{Finally, we note again that the method used here was motivated by simulations of modified gravity \cite{Li2013SimulatingMeshes}, a subject that is originally unrelated to density reconstruction. The optimal mass transportation problem, that is closely linked to the PDE we solve here, has applications in various branches of physics (e.g., nonlinear diffusion), engineering (e.g., atmosphere and ocean dynamics, aerodynamic resistance, shape and material design), mathematics (e.g., geometry, nonlinear partial differential equations), biology (e.g., leaf growth) and economics (e.g., supply-demand equilibration, structure of cities, profit maximisation, social welfare distribution) -- we hope that there are more places to find applications of this new method.}

\appendix
\section{{Detailed expressions of numerical stencils}}
\label{sec:appendixa}

{In this Appendix we present the more complicated expressions for the discretised quantities used in the code. These are for completeness only.}

{\begin{widetext}
	\begin{eqnarray}
  \bar{\nabla}^l\bar{\nabla}_m\theta\bar{\nabla}^m\bar{\nabla}_l\theta&=& ~~~\frac{1}{9h^4}\left(\theta_{i+1,j,k}+\theta_{i-1,j,k}+\theta_{i,j+1,k}+\theta_{i,j-1,k}-2\theta_{i,j,k+1}-2\theta_{i,j,k-1} \right)^2\nonumber\\
  & & +\frac{1}{9h^4}\left(\theta_{i,j+1,k}+\theta_{i,j-1,k}+\theta_{i,j,k+1}+\theta_{i,j,k-1}-2\theta_{i+1,j,k}-2\theta_{i-1,j,k} \right)^2\nonumber\\
  & & +\frac{1}{9h^4}\left(\theta_{i,j,k+1}+\theta_{i,j,k-1}+\theta_{i+1,j,k}+\theta_{i-1,j,k}-2\theta_{i,j+1,k}-2\theta_{i,j-1,k} \right)^2\nonumber\\
  & & +\frac{1}{8h^4}\left(\theta_{i+1,j+1,k}+\theta_{i-1,j-1,k}-\theta_{i+1,j-1,k}-\theta_{i-1,j+1,k}\right)^2\nonumber\\
  & & +\frac{1}{8h^4}\left(\theta_{i+1,j,k+1}+\theta_{i-1,j,k-1}-\theta_{i+1,j,k-1}-\theta_{i-1,j,k+1}\right)^2\nonumber\\
  & & +\frac{1}{8h^4}\left(\theta_{i,j+1,k+1}+\theta_{i,j-1,k-1}-\theta_{i,j-1,k+1}-\theta_{i,j+1,k-1}\right)^2+o(h^2).
    \end{eqnarray}
    \begin{eqnarray}
  & &\bar{\nabla}^l\bar{\nabla}_m\theta\bar{\nabla}^m\bar{\nabla}_n\theta\bar{\nabla}^n\bar{\nabla}_l\theta \nonumber\\
  &=& ~~~\frac{1}{9h^6}\left(\theta_{i+1,j,k}+\theta_{i-1,j,k} \right)\left[2\left(\theta_{i+1,j,k}+\theta_{i-1,j,k}\right)^2+\left(\theta_{i,j+1,k}+\theta_{i,j-1,k}\right)^2+\left(\theta_{i,j,k+1}+\theta_{i,j,k-1} \right)^2\right]\nonumber\\
  & & +\frac{1}{9h^6}\left(\theta_{i,j+1,k}+\theta_{i,j-1,k} \right)\left[2\left(\theta_{i,j+1,k}+\theta_{i,j-1,k}\right)^2+\left(\theta_{i,j,k+1}+\theta_{i,j,k-1}\right)^2+\left(\theta_{i+1,j,k}+\theta_{i-1,j,k} \right)^2\right]\nonumber\\
  & & +\frac{1}{9h^6}\left(\theta_{i,j,k+1}+\theta_{i,j,k-1} \right)\left[2\left(\theta_{i,j,k+1}+\theta_{i,j,k-1}\right)^2+\left(\theta_{i+1,j,k}+\theta_{i-1,j,k}\right)^2+\left(\theta_{i,j+1,k}+\theta_{i,j-1,k} \right)^2\right]\nonumber\\
  & & -\frac{2}{9h^6}\left(\theta_{i+1,j,k}+\theta_{i-1,j,k} \right)^2\left(\theta_{i,j+1,k}+\theta_{i,j-1,k}+\theta_{i,j,k+1}+\theta_{i,j,k-1}\right)\nonumber\\
  & & -\frac{2}{9h^6}\left(\theta_{i,j+1,k}+\theta_{i,j-1,k} \right)^2\left(\theta_{i,j,k+1}+\theta_{i,j,k-1}+\theta_{i+1,j,k}+\theta_{i-1,j,k}\right)\nonumber\\
  & & -\frac{2}{9h^6}\left(\theta_{i,j,k+1}+\theta_{i,j,k-1} \right)^2\left(\theta_{i+1,j,k}+\theta_{i-1,j,k}+\theta_{i,j+1,k}+\theta_{i,j-1,k}\right)\nonumber\\
  & & -\frac{2}{9h^6}\left(\theta_{i+1,j,k}+\theta_{i-1,j,k} \right)\left(\theta_{i,j+1,k}+\theta_{i,j-1,k}-\theta_{i,j,k+1}-\theta_{i,j,k-1}\right)^2\nonumber\\
  & & -\frac{2}{9h^6}\left(\theta_{i,j+1,k}+\theta_{i,j-1,k} \right)\left(\theta_{i,j,k+1}+\theta_{i,j,k-1}-\theta_{i+1,j,k}-\theta_{i-1,j,k}\right)^2\nonumber\\
  & & -\frac{2}{9h^6}\left(\theta_{i,j,k+1}+\theta_{i,j,k-1} \right)\left(\theta_{i+1,j,k}+\theta_{i-1,j,k}-\theta_{i,j+1,k}-\theta_{i,j-1,k}\right)^2\nonumber\\
  & & +\frac{1}{16h^6}\left(\theta_{i+1,j,k}+\theta_{i-1,j,k}+\theta_{i,j+1,k}+\theta_{i,j-1,k}-2\theta_{i,j,k+1}-2\theta_{i,j,k-1} \right)\nonumber\\
  & & ~~~~~~~~\times\left(\theta_{i+1,j+1,k}+\theta_{i-1,j-1,k}-\theta_{i+1,j-1,k}-\theta_{i-1,j+1,k}\right)^2\nonumber\\
  & & +\frac{1}{16h^6}\left(\theta_{i,j,k+1}+\theta_{i,j,k-1}+\theta_{i+1,j,k}+\theta_{i-1,j,k}-2\theta_{i,j+1,k}-2\theta_{i,j-1,k} \right)\nonumber\\
  & & ~~~~~~~~\times\left(\theta_{i+1,j,k+1}+\theta_{i-1,j,k-1}-\theta_{i+1,j,k-1}-\theta_{i-1,j,k+1}\right)^2\nonumber\\
  & & +\frac{1}{16h^6}\left(\theta_{i,j+1,k}+\theta_{i,j-1,k}+\theta_{i,j,k+1}+\theta_{i,j,k-1}-2\theta_{i+1,j,k}-2\theta_{i-1,j,k} \right)\nonumber\\
  & & ~~~~~~~~\times\left(\theta_{i+1,j+1,k}+\theta_{i-1,j-1,k}-\theta_{i+1,j-1,k}-\theta_{i-1,j+1,k}\right)^2\nonumber\\
  & & +\frac{3}{32h^6}\left(\theta_{i+1,j+1,k}+\theta_{i-1,j-1,k}-\theta_{i+1,j-1,k}-\theta_{i-1,j+1,k}\right)\nonumber\\
  & & ~~~~~~~~\times\left(\theta_{i+1,j,k+1}+\theta_{i-1,j,k-1}-\theta_{i+1,j,k-1}-\theta_{i-1,j,k+1}\right)\nonumber\\
  & & ~~~~~~~~\times\left(\theta_{i,j+1,k+1}+\theta_{i,j-1,k-1}-\theta_{i,j-1,k+1}-\theta_{i,j+1,k-1}\right)+o(h^2).
    \end{eqnarray}
\end{widetext}
}

\

\begin{acknowledgements}
We thank Christian Arnold, Jianhua He, Wojciech Hellwing, Mark Neyrinck, Rien van de Weygaert, {Xin Wang and Hongming Zhu for useful discussions and comments, and Xin Wang for provide the simulations used in making the BAO test}. YS is financially supported by University of Science and Technology of China, and thanks the host by the Institute for Computational Cosmology (ICC) at Durham University when the work described in this paper was carried out. MC is supported by STFC Consolidated Grant ST/L00075X/1. BL is supported by the European Research Council (ERC-StG-716532-PUNCA) and STFC Consolidated Grants ST/P000541/1 \& ST/L00075X/1. The work described here used the DiRAC Data Centric system at Durham University, operated by the ICC on behalf of the STFC DiRAC HPC Facility (www.dirac.ac.uk). This equipment was funded by BIS National E-infrastructure capital grant ST/K00042X/1, STFC capital grants ST/H008519/1 and ST/K00087X/1, STFC DiRAC Operations grant ST/K003267/1 and Durham University. DiRAC is part of the National E-Infrastructure.
\end{acknowledgements}

\bibliography{recon.bib}

\end{document}